\documentclass[prd,twocolumn,nofootinbib,showpacs,superscriptaddress]{revtex4}

\usepackage{graphicx}

\newcommand{\beq}{\begin{equation}} \newcommand{\eeq}{\end{equation}}
\newcommand{\bea}{\begin{eqnarray}} \newcommand{\eea}{\end{eqnarray}}

\newcommand{\mg}{M_{{\rm GUT}} } \newcommand{\mpl}{M_{{\rm Planck}} }
\newcommand{\ut}{U_T} 

\newcommand{\beqa}{\begin{eqnarray}}
  \newcommand{\eeqa}{\end{eqnarray}}

\begin{document}


\title{Light GUT Triplets and Yukawa Splitting}

\author{Subhendu Rakshit}
\email{srakshit@physics.technion.ac.il}
\affiliation{Physics Department, Technion, Haifa 32000, Israel}

\author{Guy Raz}
\email{guy.raz@weizmann.ac.il}
\affiliation{Particle Physics Department, Weizmann Institute of
  science, Rehovot 76100, Israel}

\author{Sourov Roy}
\email{roy@pcu.helsinki.fi}
\affiliation{Helsinki Institute of Physics, P.O.Box 64, FIN-00014 University of Helsinki, Finland}

\author{Yael Shadmi}
\email{yshadmi@physics.technion.ac.il}
\affiliation{Physics Department, Technion, Haifa 32000, Israel}


\begin{abstract}
  Triplet-mediated proton decay in Grand Unified Theories (GUTs) is
  usually suppressed by arranging a large triplet mass.  Here we
  explore instead a mechanism for suppressing the {\sl couplings} of
  the triplets to the first and second generations compared to the
  Yukawa couplings, so that the triplets can be light.  This mechanism
  is based on a ``triplet symmetry'' in the context of product-group
  GUTs.  We study two possibilities.  The first possibility, which
  requires the top Yukawa to arise from a non-renormalizable operator
  at the GUT scale, is that all triplet couplings to matter are
  negligible, so that the triplets can be at the weak scale, giving
  new evidence for grand unification. The second possibility is that
  some triplet couplings, and in particular $T t b$ and $T \bar{t}
  \bar{l}$, are equal to the corresponding Yukawa couplings.  This
  would give a distinct signature of grand unification if the triplets
  were sufficiently light.  However, we derive a model-independent
  bound on the triplet mass in this case, which is at least 10$^6$GeV.
  Finally, we construct an explicit viable GUT model based on Yukawa
  splitting, with the triplets at 10$^{14}$ GeV, as required for
  coupling unification to work.  This model requires no additional
  thresholds below the GUT scale.
\end{abstract}

\maketitle

\section{Introduction.}

One of the main challenges of supersymmetric Grand Unified Theories
(GUTs), is the rate of proton decay mediated by the GUT partners of
the Standard-Model Higgs doublets.  These color-triplet partners
couple to Standard Model fermions violating both baryon and lepton
number, through \beq\label{tripcoupl} W= y^{Tqq}_{ij} T q_i q_j +
y^{T\bar{u}\bar{e}}_{ij} T \bar{u}_i \bar{e}_j + y^{\bar{T}q l}_{ij}
\bar{T} q_i l_j + y^{\bar{T}\bar{u}\bar{d} }_{ij} \bar{T} \bar{u}_i
\bar{d}_j \ .  \eeq Here $T$ and $\bar{T}$ are the color-triplets,
barred fields are SU(2) singlets and unbarred fields are SU(2)
doublets, and $i,j$ are generation indices.  Typically, the triplet
couplings of~(\ref{tripcoupl}) and the Yukawa couplings $y^U_{ij}
H_Uq_i \bar{u}_j + y^D_{ij} H_D q_i \bar{u}_j$, originate from the
same GUT superpotential terms, so that $y^{Tqq}=y^{T\bar{u}\bar{e}}=
y^U$ and $y^{\bar{T}q l}=y^{\bar{T}\bar{u}\bar{d}}= y^D$.  These
couplings mediate proton decay at the level of dimension-five
operators.  To suppress this contribution, one typically tries to
arrange a GUT-scale triplet mass, while keeping the Standard Model
Higgses at the weak scale.  In fact, even models with triplets at the
GUT scale are in conflict with current experimental bounds on the
proton lifetime.  For example, in minimal SU(5), the lower bound on
the triplet mass is about
10$^{17}$GeV~\cite{goto,hisano,murayama,ecw}.

Here we explore instead the possibility that triplet {\sl couplings}
to matter are smaller than the Yukawas, namely, $y^{Tqq},
y^{T\bar{u}\bar{e}} \ll y^U$ and $y^{\bar{T}q l},
y^{\bar{T}\bar{u}\bar{d}}\ll y^D$.  This relaxes the proton decay
bound on the triplet mass, allowing the triplets to be lighter than
the GUT scale.

Thinking about triplets below the GUT scale is motivated by two
reasons.  The observation of proton decay, taken together with
coupling unification, would be a strong indication for GUTs but still
far from conclusive evidence.  It is therefore intriguing to see
whether the triplets can be sufficiently light that they can give a
more direct experimental signature of the GUT.  We will exhibit a
model in which all triplet couplings to matter are tiny, so that the
triplets can be around the weak scale.  To allow for coupling
unification, we will need to arrange an extra pair of light doublets.
Thus, at low energies, there would be new fields in complete $SU(5)$
representations, providing additional evidence for grand unification.

An even more spectacular GUT signature would potentially come from
models in which triplet couplings to first and second generation
fields are suppressed, so that the proton does not decay too fast, but
some triplet couplings to third generation fields are order one.  This
is naturally the case in many of our models.  However, we will show
that just the presence of the couplings $y^{Tqq}_{33}$ and
$y^{T\bar{u}\bar{e}}_{33}$ requires the triplets to be heavier than at
least 10$^6$GeV, in order to satisfy the bounds on proton decay.

As mentioned above, current bounds on the proton lifetime imply a
lower bound of 10$^{17}$GeV on the triplet mass in minimal
SU(5)~\cite{murayama}.  On the other hand, for coupling unification to
occur in minimal SU(5), the triplet mass should be close to
10$^{14}$GeV~\cite{murayama}.  This mismatch is the second motivation
for triplets below the unification scale~\cite{senj}.  We will present
a simple model with triplets at 10$^{14}$GeV.  In this model, all
$\bar{T}$ couplings to matter are very small, so that the
dimension-five contribution to proton decay is suppressed, and the
dominant contribution is from $X$ and $Y$ gauge boson exchange.  Many
more variants with these properties can be constructed.

The possibility that triplet couplings to matter are small was also
discussed in~\cite{ramond}-\cite{haba}.  In~\cite{ramond}, fermion
masses were assumed to originate from two different Higgs fields, such
that the two triplet contributions to proton decay cancel.
Ref.~\cite{dvali} considered an $SO(10)$ GUT, and argued that the
triplet couplings are forbidden by a symmetry.  It is therefore
closest in spirit to our current work.  It was assumed, however, that
the top Yukawa originates from a higher dimension operator suppressed
by $\mg$ only, and not by a higher scale. Thus, there is no energy
region in which there exists a sensible effective theory with all
couplings being of the same order of magnitude.  The mechanism of
Ref.~\cite{haba} involves an extra dimension.

Our models are all based on a ``triplet symmetry'', $U_T$, that
distinguishes between the triplets and the doublets~\cite{witten}.  We
take $U_T$ to be either a $U(1)$ or a $Z_N$.  Since it does not
commute with the GUT group, $U_T$ must arise from the combination of
some global symmetry and a subgroup of the GUT.  Moreover, in order
for some coupling involving the triplet to have a different $U_T$
charge from the corresponding doublet coupling, the GUT group must be
semi-simple~\cite{witten}.  For concreteness, we will take the GUT to
be $SU(5)\times SU(5)$.  Such a setup was used
in~\cite{Barbieri:1994jq,Barr:1996kp, witten, dns} to generate a
doublet-triplet mass hierarchy.  As was pointed out in~\cite{witten},
the symmetry may also lead to a suppression of dimension-five
operators contributing to proton decay, through the suppression of
triplet coupling to fermions.  In~\cite{dns}, explicit models that
have these properties were constructed. However, the possibility of
exploiting ``Yukawa splitting'' in order to lower the triplet mass was
not explored\footnote{In the models of~\cite{babubarr}, the Dirac mass
  for the triplets was suppressed, resulting in smaller dimension-five
  operators.}.

As we will see, a triplet-matter coupling can be different from the
corresponding Yukawa coupling whenever the relevant Higgs field and
matter fields transform under different $SU(5)$ factors.  Then, the
Yukawa coupling must arise from a non-renormalizable term in the GUT
superpotential, which involves some combination of GUT-breaking
fields.  The key point is that this operator does not lead to a
triplet coupling: the triplet coupling can only come from a different
GUT term, involving a different combination of GUT-breaking fields.
If some Yukawa coupling has its origin in a non-renormalizable GUT
term, it involves some power of $\mg/\mpl$.  Thus, if the Planck scale
is 10$^{18}$GeV, the top Yukawa must arise from a renormalizable GUT
term, and $y^{Tqq}_{33}\sim y^{T\bar{u}\bar{e}}_{33}\sim1$.  We
therefore distinguish between two classes of models.  {\bf a.} The top
Yukawa coupling is renormalizable at the GUT scale.  Then,
$y^{Tqq}_{33}\sim y^{T\bar{u}\bar{e}}_{33}\sim1$.  {\bf b.} The top
Yukawa originates from a non-renormalizable term at the GUT scale. In
this case, all triplet couplings to matter can be small.  This is only
possible if the Planck scale is around 10$^{17}$GeV, as would be the
case with a small extra dimension.  Then, the top Yukawa can be of the order
of ${\cal O}(10^{-1})$
at the GUT scale, with running effects driving it to order one at
the weak scale.

\section{Models: basic structure.}

We now turn to the basic structure of our models,
following~\cite{dns}.  The models have the symmetry $SU(5)_1\times
SU(5)_2\times U_T^0$, where $U_T^0$ is a global symmetry which we will
take to be either a $U(1)$ or a $Z_N$.  The symmetry is broken by two
sets of bifundamental fields $\Phi_3\sim(5,\bar5,1)$,
$\bar\Phi_3\sim(\bar5,5,-1)$, $\Phi_2\sim(5,\bar5,q)$, and
$\bar{\Phi}_2\sim(\bar5,5,-q)$, where the third entry corresponds to
the $U_T^0$ charge.  For $U_T^0=Z_N$ ($U_T^0=U(1)$) we take
$q={N-3\over2}$ ($q=-{3\over2}$).

For the VEVs \beq\label{vevs}
\begin{array}{l}
\langle\Phi_3\rangle=\langle\bar{\Phi}_3\rangle=
{\rm diag}(v_3,v_3,v_3,0,0), \smallskip \\
\langle\Phi_2\rangle=\langle\bar{\Phi}_2\rangle= {\rm diag}(0,0,0,v_2,v_2)\ ,
\end{array}
\eeq with $v_2\sim v_3\sim 10^{16}$GeV, the symmetry is broken to
$[SU(3)\times SU(2)\times U(1)]_{{\rm SM}}\times\ut$, where $\ut$ is a
combination of $U_T^0$ and a discrete hypercharge subgroup of
$SU(5)_1$.  The standard-model gauge group lies in the diagonal $SU(5)$, so that
the standard-model gauge couplings all start from the diagonal $SU(5)$ coupling,
and hypercharge is quantized as in minimal $SU(5)$.

As was shown in~\cite{dns}, one can add three $SU(5)_1$ adjoints and a
singlet such that the direction~(\ref{vevs}) is flat, all uneaten
GUT-breaking fields get heavy, and the ratio $v_3/v_2$ is naturally of
order one.

There are now different possible choices for the MSSM matter fields,
since they can transform under either of the two $SU(5)$ gauge groups.
This, and the charge assignments under the triplet symmetry, will
define the different models we consider.

\section{Non-renormalizable top Yukawa.}
\label{non}

As mentioned above, if the Planck scale is near 10$^{17}$GeV, the top
Yukawa may originate from a non-renormalizable term at the GUT scale.
It is then easy to construct models in which the triplets can be as
light as the weak scale. Take the Standard Model Higgs fields to come
from $h\sim(5,1,q)$ and $\bar{h}\sim(\bar{5},1,0)$ under
$SU(5)_1\times SU(5)_2\times U_T^0$, and the matter fields to be three
copies of $(1,10,0)$ and $(1,\bar{5},-q)$.  We also add the fields
$N\sim(5,1,2)$, $\bar{N}\sim(1,\bar{5},-1)$, $\bar{M}\sim(\bar5,1,1)$,
$M\sim(1,5,2)$.  $N$ and $\bar{N}$ are needed to restore coupling
unification, and $M$ and $\bar{M}$ are required for anomaly
cancellation.

The Yukawa couplings now come from \beq\label{wnew} {1\over\mpl}\,
\bar\Phi_2 h (1,10) (1,10) + {1\over\mpl}\, \Phi_2 \bar{h} (1,10)
(1,\bar5) \ , \eeq where we suppress generation indices.  All Yukawa
couplings are suppressed by $\mg/\mpl\sim10^{-1}$ at the GUT scale.  
Since running to the weak scale enhances the top Yukawa coupling by
roughly 3, these models are viable if the order-one coefficient multiplying 
the top coupling at the GUT scale is around 3.

As long as the triplet symmetry is unbroken, the analogous terms with 
$\Phi_3$ and $\bar\Phi_3$ are forbidden, and the triplets have no couplings 
to matter.
In addition, the triplet symmetry forbids both a triplet mass and a $\mu$ term.
Thus, it should ultimately be broken in order to generate a $\mu$ term.  
The most attractive possibility  is that this breaking is related to 
supersymmetry breaking. 
Then some triplet-matter couplings would typically be generated, suppressed by
powers of the weak scale over the Planck scale.

Since the triplet Higgses are light, coupling unification is lost and
we must split another GUT multiplet to restore it.  It is easy to do
so by allowing the superpotential term $N\bar\Phi_3\bar{N}$.  The
triplets in $N$ and $\bar{N}$ get a $\mg$ mass, but their doublet
partners remain light.

As mentioned above, the fields $M$ and $\bar{M}$ are only needed for
anomaly cancellation. The triplets and doublets of these fields cannot
both get mass at the GUT scale because of the triplet symmetry.  The
simplest possibility is then to forbid their masses altogether in the
limit of unbroken supersymmetry.  Indeed, with the $U_T$ charges
specified above, no masses are allowed for $M$ and $\bar{M}$.  Once
the triplet symmetry is broken, the triplets and doublets in $M$ and
$\bar{M}$, as well as the doublets in $N$ and $\bar{N}$, will get mass
around the supersymmetry-breaking scale.\footnote{We have only shown
  here one possible choice of $U_T^0$ charges.  Moreover, there are
  other possibilities for forbidding mass terms for $M$ and $\bar{M}$
  apart from using the triplet symmetry.  For example, we can impose
  an R-symmetry symmetry instead.  Then again, the triplets and
  doublets in $M$ and $\bar{M}$ will get mass near the supersymmetry
  breaking scale.}

At low energies, we then have two extra $5$'s and two extra $\bar5$'s: 
One pair coming from Higgs triplets and the doublets of $N$ and $\bar{N}$, 
and the other from $M$ and $\bar{M}$. This would provide an additional hint
for a grand-unified structure. However, because the light triplets
have no couplings to the matter fields, it is impossible to tell that
they are actually the GUT-partners of the Higgs doublets.

\section{Renormalizable top Yukawa.}

We now turn to models in which the top Yukawa comes from a
renormalizable term at the GUT scale, so that $y^{Tqq}_{33}\sim
y^{T\bar{u}\bar{e}}_{33}={\cal O}(1)$.  Below the triplet mass scale, the
theory contains four-fermion operators involving the top quark, which
might be probed by future experiments if the triplet mass $M_T$ were
low enough.  We will first derive a model-independent bound on the the
triplet mass in this case. In order to do that we 
will consider the ``most favorable'' scenario, that is, we will
assume that all other $T$ and $\bar{T}$ couplings vanish at the tree-level. 
We emphasize that such an assumption is unnatural. 
Terms that vanish at tree level would
appear at one loop as we demonstrate shortly. 
Moreover, it is probably
impossible to forbid all $T$ couplings in~(\ref{tripcoupl}) for
$i,j\neq3$, while generating acceptable fermion masses, and in
particular, non-zero mixing of the third generation with the first
two.  Still, this unnatural scenario will allow us to obtain a useful
bound for phenomenological purposes, because even in this most
favorable case, the resulting bound is very strong. 

We assume then that the only nonzero $T$ couplings are $y^{Tqq}_{33}$
and $y^{T\bar{u}\bar{e}}_{33}$.  Upon rotating to the mass basis, tree
level $T$ couplings to first generation fields will typically be
generated.  These are however model dependent.  For example, if the up
mass matrix is diagonal, such terms do not appear.

\begin{figure}
  \includegraphics[width=0.9\linewidth]{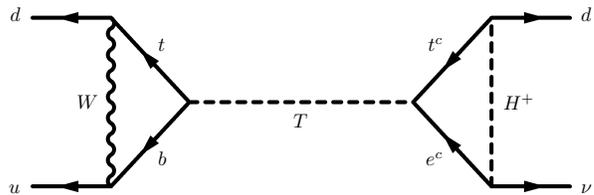}
\caption{\label{fig1} Two-loop $d=6$ diagram for proton decay}
\end{figure}

One can still obtain a model-independent bound from loop diagrams,
such as the two-loop diagram of Fig.\ref{fig1}.
Assuming that, in a basis with well-defined $U_T$ charges, the up
matrix is diagonal\footnote{Otherwise, proton decay may be generated
  already at tree level.}, the $W$-loop is proportional to
\beq\label{leftloop} {\alpha_w\over4\pi}\, V_{td}^* \sum_i V_{ti}^*
V_{ui} {m_{d_i}^2\over m_t^2-m_W^2} \ln\left({m_t^2\over m_W^2}\right)
\sim 10^{-10}\ , \eeq where $V$ is the CKM matrix.  The charged Higgs
loop scales as \beq\label{rightloop} {\alpha_w\over4\pi}\, V_{td}^* \,
{m_l\over m_W}\sim \cases{ 10^{-6} & $l=\tau$ \cr 10^{-10} & $l=e$ }\
.  \eeq 
Thus we see that single $T$ exchange, even in the presence of
triplet couplings to third-generation quarks only, implies a lower
bound on the triplet mass between 10$^6$ and 10$^8$GeV, depending on
whether the lepton belonging to the same 10 as the top is the electron
or the tau.  It is therefore unlikely that the triplets can be
detected in any foreseeable experiment.

In fact, as was mentioned above, in any natural model one would expect
to have some other nonzero triplet couplings besides the ones we
considered here, so that proton decay would occur already at tree
level.  If the model is based on a symmetry which is broken by some
small parameter, it is reasonable to expect the tree contribution and
the loop contribution to involve the same parametric suppression.
Thus, in any natural model, the bound on the triplet mass would be
about one loop factor, or two orders of magnitude, stronger than what
we found above.

Indeed, the most promising models from the point of view of getting
small triplet couplings to the first generations, are models with the
Higgses in the first SU(5), one 10 in the first SU(5) (so that the top
Yukawa is renormalizable), and all other matter fields in the second
SU(5), but we have been unable to find any model of this type for
which proton decay would allow triplets below, roughly, 10$^{11}$GeV.

\section{Triplets at 10$^{14}$GeV.}

We will now use ``Yukawa splitting'' to obtain a model which
reconciles the conflicting requirements from proton decay and coupling
unification on the triplet mass.  Many models of this type can be
constructed, but we will present one simple example.

We take the Higgses to be $h=(5,1,0)$ and ${\bar h}=(1,\bar5,0)$, and
the matter fields to be three copies of $(10,1,0)$ and $(\bar5,1,q)$
with $q={N-3\over2}$ ($-3/2$) for $Z_N$ ($U(1)$)\footnote{This choice
  was also mentioned in~\cite{witten}.  There however, the triplets
  got mass at the GUT scale.  }.  To cancel anomalies, we also add the
fields $\bar{M}\sim(\bar5,1,1)$ and $M\sim(1,5,1)$.

As in the usual case, the up-type Yukawas originate from the
superpotential terms $h (10,1) (10,1)$, so that
$y^{Tqq}=y^{T\bar{u}\bar{e}}=y^U$.  On the other hand, down-type
Yukawas must come from non-renormalizable terms involving a
bifundamental, \beq
\label{down}
y^D_{ij}\, {1\over\mpl}\, \bar\Phi_2 \bar{h} (10,1)_i\,(\bar5,1)_j\ ,
\eeq so that all down Yukawas are uniformly suppressed by
$\langle\bar\Phi_2\rangle/\mpl\sim 10^{-2}$.  The
superpotential~(\ref{down}) does {\sl not} give rise to a
$\bar{T}$-matter coupling, since that requires a $\bar\Phi_3$ instead
of $\bar\Phi_2$ in eqn.~(\ref{down}).

The dominant triplet contribution to proton decay now comes from
single $T$ exchange. This is very similar to the usual $X$ and $Y$
gauge-boson mediated decay, but is much smaller, since the couplings
involved are the Yukawas rather than gauge couplings.  We then find
that the bound on the triplet mass implied by proton decay is roughly
$10^{12}$GeV.

The triplet symmetry so far forbids a triplet mass as well as a $\mu$
term.  In order to generate an acceptable triplet mass, we assume that
the symmetry is broken by some small parameter $\eta$ (which arises,
say, as $\langle S\rangle\over\mpl$ with $S$ a fundamental or
composite gauge singlet).  For $\eta$ of charge $+1$, with $\eta\sim
10^{-2}$, the superpotential term \beq\label{eta} \eta\, h \bar\Phi_3
\bar{h}\ , \eeq is allowed and gives a 10$^{14}$GeV triplet mass.
Coupling unification is then recovered, while proton decay is below
experimental bounds.

The remaining aspect of ``doublet-triplet'' splitting, the $\mu$
problem, is readily solved as well.  Suppose first that the triplet
symmetry is a $U(1)$.  Then, once we allow the coupling~(\ref{eta}),
the $\mu$ term is automatically forbidden by the triplet symmetry and
holomorphy~\cite{nir}.  One can alternatively generate an acceptable
$\mu$ term by taking the triplet symmetry to be discrete~\cite{lns}.
For example, with $U_T=Z_{17}$ we have $M_T\sim10^{14}$~GeV as before
and $\mu\sim \eta^7\mg\sim 10^2$~GeV.

Since the triplet symmetry is now broken, $\bar{T}$ couplings to
matter may in principle be generated from terms such as $\eta^p
\bar\Phi_3\bar{h} (10,1) (\bar5,1)$.  One can check that, in the
examples we consider, these terms are either forbidden, or are very
suppressed, so that the dominant triplet contribution to proton decay
is still from single $T$ exchange as described above.

One could also worry that, once the triplet symmetry is broken, the
pattern of VEVs~(\ref{vevs}) is no longer protected, and some triplet
couplings are re-introduced from terms already present in the
superpotential, such as~(\ref{down}).  However, the
pattern~(\ref{vevs}) can be guaranteed using additional symmetries.
One possibility is an $R$ symmetry under which the bifundamentals and
$\eta$ have zero charges, and so cannot appear alone in the
superpotential.  This $R$ symmetry was present anyway in the models
of~\cite{dns} in order to ensure a flat potential for the
bifundamentals.

Finally we comment on the fields $M$ and $\bar{M}$. 
Just as the analogous fields $M$ and $\bar{M}$ of the model
of section~\ref{non}, their mass is protected by the triplet symmetry
and supersymmetry. 
Alternatively, their masses can be forbidden by the $R$-symmetry
discussed above.
In any case, once supersymmetry is broken, these fields will get mass
near the supersymmetry-breaking scale.

To summarize, this model avoids the mismatch of coupling unification
by having the triplets near 10$^{14}$~GeV, and gives proton decay
below current experimental bounds.  At low energies, there are new
fields in complete $SU(5)$ representations, giving additional evidence
for grand unification.

\section{Concluding Remarks.}

In the examples we discussed, all matter fields transform under a
single SU(5).  More generally, if either the 10s or the $\bar5$s are
split between the different SU(5) factors, the triplet symmetry will
necessarily behave as a horizontal symmetry: Some matter fields will
have generation-dependent $U_T$ charges.  Clearly, these models give
interesting patterns of fermion masses, which can significantly differ
from the usual GUT relations.

We have assumed that there are no flavor-violating contributions from
the superpartner sector.  This holds for example for universal scalar
masses.

Finally, we stress that, even in minimal GUT models,
the triplet contribution to proton decay is
highly sensitive to the details of the Yukawa couplings.
For example, if the up-quark mass vanishes, so that, in the up-mass
basis, $y^{u}_{11}=0$, there is no dimension-five triplet
contribution to proton decay.

To summarize, we have shown that triplets below the
GUT scale can be compatible with proton stability. If all
triplet-matter couplings are suppressed, the triplets can be near the
weak scale, and thus directly detectable.  
This would provide additional evidence for grand unification,
which would otherwise be very hard to establish conclusively,
even if proton decay is observed one day.  

If the triplets only have
large couplings to the top, their mass must be at least 10$^6$GeV. It
would therefore be impossible to detect them.  However, it is still
interesting to consider triplets near 10$^{14}$GeV, since then they
supply just the right contributions to the running so that MSSM gauge
couplings do unify.  We present simple models that realize this
possibility, with no other thresholds below the GUT scale.

\acknowledgments

We thank Yuval Grossman, Yossi Nir, Yuri Shirman, and Pierre
Ramond for useful discussions.  We also thank Pierre Ramond for
pointing out ref.~\cite{ramond}, and David~E.~Kaplan for pointing
out ref.~\cite{dvali} to us.  S.~Rakshit thanks Fermilab and SLAC,
and Y.S. thanks LANL, The Aspen Center for Physics, SLAC and
UCSC-SCIPP for hospitality while this work was being completed.
The research of S.~Roy was supported by a Lady Davis grant.
The research of Y.S. is supported in part by the Israel Science 
Foundation (ISF) under grant 29/03, 
and by the United States-Israel Binational Science 
Foundation (BSF) under grant 2002020.

\end{document}